\documentclass{article}
\usepackage[T1]{fontenc} 
\usepackage[utf8]{inputenc} 
\usepackage{ismir,amsmath,cite,url}
\usepackage{graphicx}
\usepackage{color}

\def\lbd{}	

\newcommand{\iec}{i.\,e.,\,}

\newcommand{\egc}{e.\,g.,\,}

\newcommand{\Fig}[1]{Figure~\ref{#1}}

\usepackage{lineno}
\usepackage[bookmarks=false]{hyperref}

\title{Fully Automatic Page Turning on Real Scores}

\multauthor
{Florian Henkel$^1$ \hspace{1cm} Stephanie Schwaiger \hspace{1cm} Gerhard Widmer$^{1, 2}$} { 
 $^1$ Institute of Computational Perception, Johannes Kepler University, Linz, Austria\\
$^2$ LIT Artificial Intelligence Lab, Linz Institute of Technology, Austria \\
{\tt\small florian.henkel@jku.at}
}

\def\authorname{F. Henkel, S. Schwaiger, and G. Widmer}

\begin{document}

\maketitle
\begin{abstract}

We present a prototype of an automatic page turning system that works directly on real scores, i.e., sheet images, without any symbolic representation.
Our system is based on a multi-modal neural network architecture that observes a complete sheet image page as input, listens to an incoming musical performance, and predicts the corresponding position in the image.
Using the position estimation of our system, we use a simple heuristic to trigger a page turning event once a certain location within the sheet image is reached.
As a proof of concept we further combine our system with an actual machine that will physically turn the page on command.

\end{abstract}
\section{Introduction}\label{sec:introduction}

Classical piano pieces are usually too long to fit a single page of sheet music. This leads to the problem that a musician will have to turn the page once its end is reached.
As this requires the performer to move her hands away from the instrument, this is cumbersome and sometimes strictly impossible without pausing or making errors. 
During live performances pianists will often have a human page turner standing aside, who reads along in the score and turns the page when required.

In \cite{ArztWD08_PageTurning_ECAI}, the authors propose an approach that automatically performs this task using dynamic time warping (DTW). However, their system relies on a MIDI representation of the score (which must be annotated with page endings or appropriate page turning points) that is first synthesized to audio and subsequently aligned to the incoming musical performance \cite{Arzt16_MusicTracking_PhD}. 
In the following, we present a prototype of a system that is able to turn pages fully automatically based on raw sheet images as input, \iec no symbolic score representation or manual pre-processing is required.\footnote{Code and demo videos of our system will be made available on-line: \url{https://github.com/fhenkel/page_turner}.}

\section{Automatic Page Turning with Sheet Images}

The basis of our page turner is the sheet-image-based score following system presented in \cite{HenkelW21_CYOLO_EUSIPCO}. 
This system takes a complete sheet image page as input and predicts the most likely bounding box around the note region that matches the currently incoming performance audio. 
Performance and tracking accuracy are analyzed in detail in \cite{HenkelW21_CYOLO_EUSIPCO}.

Using this position estimation, we define a heuristic to turn the page once a certain location is reached in the sheet image. Specifically, we will turn it once the position is estimated to be halfway into the last system on a page.

\begin{figure}
\centering
\includegraphics[width=1.\columnwidth]{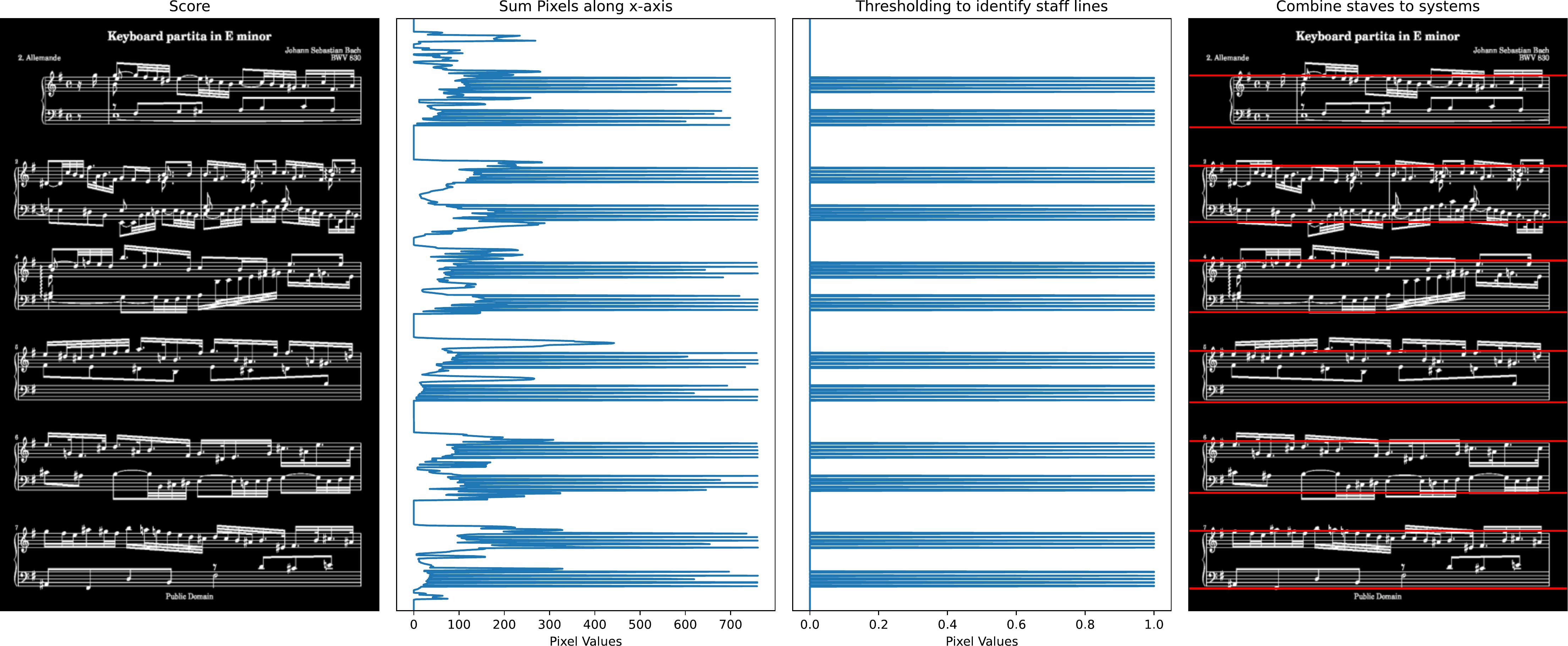}
 \caption{Score processing to detect system locations on a page:
 (a) binarized score image; (b) horizontally summing up the pixel values along the x-axis; (c) thresholding the summed pixel values to identify staff lines; (d) combining staff lines to systems (given as start and end points on the y-axis).
 }
\label{fig:processing}
\end{figure}

To that end, we first need to identify the last system on a page, which is done using simple mathematical operations (see \Fig{fig:processing}). Once we have identified the last system (depending on the quality of the input score this will only be a rough estimate), we can compare this information to our predicted position within the overall score. In case our prediction falls into the last system we further check if the tracker is past a certain location, \egc halfway in, and trigger a page turning event if desired.

\section{Live Demonstration}

\begin{figure*}[t!]
\centering
\includegraphics[width=0.8\textwidth]{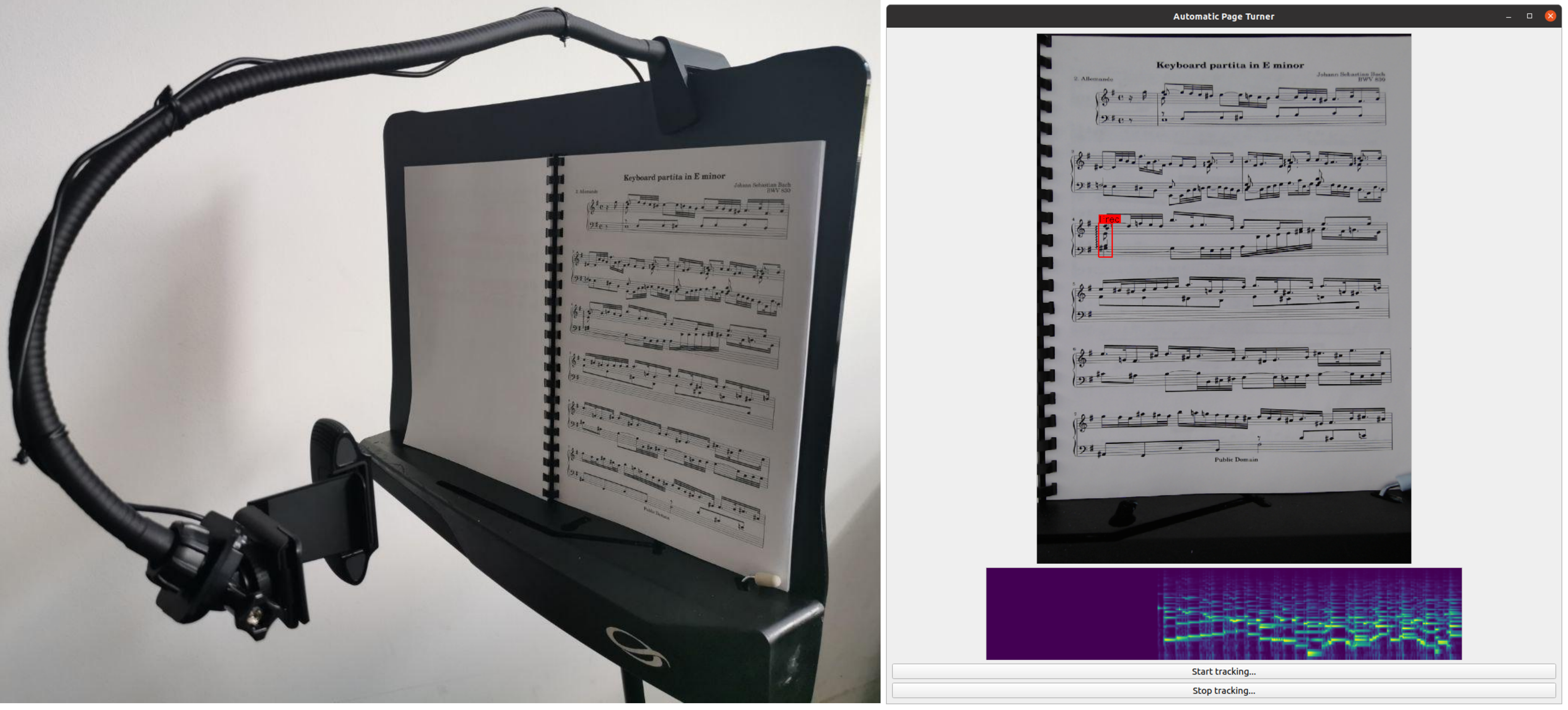}
 \caption{Page turning device equipped with a webcam (left) and the user interface of our application showing the input and prediction of our system. (right).}
\label{fig:page_turner}
\end{figure*}

For our live demonstration we connect the page turning system to the same physical device  as described in \cite{ArztWD08_PageTurning_ECAI}. This device is able to turn the pages of music scores using two mechanical fingers that first lift a page slightly, and subsequently push it to the other side.
Furthermore, we equip the device with a camera, providing our system with a live input of the music score (cf. \Fig{fig:page_turner}). Apart from the general difficulty of score following with raw sheet images, this additionally requires our system to be robust to various image distortions caused by lighting and the camera in general.

To alleviate these issues, we convert the RGB camera input to a gray scale image and further apply  adaptive thresholding with a Gaussian window.
The resulting binarized image is then fed to the score following system.

As the score follower is not restricted in its predictions, \iec it can predict a bounding box anywhere on the sheet image, we further apply several post-processing constraints to improve stability.
These constraints formalize the typical behavior of a human reading sheet music --- from left to right and top to bottom (neglecting repeats in the music).
This is done by filtering the predicted locations according to the following rules:
at the beginning of each page we start in the first system (top left);
the tracker is only allowed to jump between adjacent systems (\egc it is not possible to immediately jump from  the first to the last system); additionally, jumping backward is only allowed if the predicted location is above a certain confidence level (which we set to $0.5$).\footnote{Besides bounding box coordinates the tracker also outputs confidence scores. See \cite{HenkelW21_CYOLO_EUSIPCO} for more details.}
The underlying assumption is that the music does not contain any repeats that would require the score follower to significantly jump around in the sheet image, which is the case for the data used to train the score following system.

The videos accompanying this work show that our system works quite well with synthetically rendered audio and the provided camera input.\footnote{See supplementary videos.}
Likewise the pages are turned correctly if a real piano recording is provided instead, however the tracking is less accurate.
This indicates that the tracker is better capable of handling image than audio degradations.

Our experiences with the page turner so far suggest that it is very robust, especially if the systems on the image are large and not formatted too close together.
For pieces with small-sized systems as well as very repetitive note patterns the tracker struggles to correctly follow the performances.
Furthermore, it is important for correct page turning that the tracker is not lost within the last system on a page.

\section{Conclusion and Future Work}
We have presented a first prototype of a sheet-image-based automatic page turning system. Our system successfully turns pages given clean score sheet images and is additionally capable of handling (noisy) camera input.

For future work, there are still several things to consider. First, we do not yet handle repeats in the music, which imply jumps in the score when reading. While this should be straightforward for repeats within a single page, it poses a significant technical challenge to our score following model when the repeat spans multiple pages; the page turning device itself is mechanically incapable of turning pages back.

Second, we currently only consider a fixed location when a page should be turned (halfway into the last system on the page).
This ignores both the actual musical content of the score (which may be more or less dense) and the tempo at which it is being played.
One possibility could be to extend our current score following system to additionally predict whether the end of a page is close. For example, one could define an ``end-of-page`` (EOP) indicator which should be $1$ for every time step from, \egc one second up until the last note position on a page, and $0$ otherwise. 
In that way, the system implicitly considers the tempo of a piece, and dynamically turns pages a certain number of time steps before the end is reached.

Alternatively, one could also try to estimate a score reading tempo from the predicted positions, \iec how fast the tracker is moving within the sheet image, and use this information together with the estimated location of the last system to turn pages accordingly.
In future work, we will evaluate and compare these different strategies to determine the most suitable and robust approach.

\section{Acknowledgements}
 This work has been supported by the European Research Council (ERC) under the European Union's Horizon 2020 research and innovation program (grant agreement number 670035, project "Con Espressione"). We would like to thank Jan Haji\v{c} Jr.~and Carlos Eduardo Cancino Chac\'on for performing and recording the test pieces for us.

\bibliography{main}

\end{document}